\documentclass[%
 twocolumn, aip,
jap,
reprint,%
superscriptaddress,
longbibliography,
nourl,
noeprint
]{revtex4-1}

\usepackage{graphicx}
\usepackage{dcolumn}
\usepackage{bm}

\usepackage[utf8]{inputenc}
\usepackage[T1]{fontenc}
\usepackage{mathptmx}
\usepackage{etoolbox}
\usepackage{amsmath}

\makeatletter
\let\@fnsymbol\@fnsymbol@latex
\@booleanfalse\altaffilletter@sw
\makeatother

\begin{document}


\title[]{Electrically and optically active charge carrier traps in silicon-doped few-layer GaSe}

\author{M. Bissolo}%
\email[]{michele.bissolo@tum.de}
\affiliation{ 
Walter-Schottky-Institut and TUM School of Natural Sciences, Technische Universit\"at M\"unchen, Am Coulombwall 4, 85748 Garching, Germany
}

\author{R. Li}%
\altaffiliation{ 
Current address: Max Planck Institute for Chemical Physics of Solids, N\"othnitzer Str. 40, 01187 Dresden, Germany
}
\affiliation{ 
Walter-Schottky-Institut and TUM School of Natural Sciences, Technische Universit\"at M\"unchen, Am Coulombwall 4, 85748 Garching, Germany
}

\author{M. Ogura} 
\affiliation{
Department of Chemistry, Ludwig-Maximilians-Universit\"at M\"unchen,  Butenandtstrasse 5-13, 81377 M\"unchen, Germany
}

\author{Z. Sofer}
\affiliation{
Department of Inorganic Chemistry, University of Chemistry and Technology Prague, Technick\'a 5, 166 28 Prague 6, Czech Republic
}


\author{S. Polesya} 
\affiliation{
Department of Chemistry, Ludwig-Maximilians-Universit\"at M\"unchen,  Butenandtstrasse 5-13, 81377 M\"unchen, Germany
}

\author{D. Han} 
\affiliation{
School of Materials Science and Engineering, Jilin University, Changchun, 130012 China
}

\author{A. W. Holleitner}%
\affiliation{ 
Walter-Schottky-Institut and TUM School of Natural Sciences, Technische Universit\"at M\"unchen, Am Coulombwall 4, 85748 Garching, Germany
}

\author{C. Kastl}%
\affiliation{ 
Walter-Schottky-Institut and TUM School of Natural Sciences, Technische Universit\"at M\"unchen, Am Coulombwall 4, 85748 Garching, Germany
}

\author{G. Koblm\"uller}%
\affiliation{ 
Walter-Schottky-Institut and TUM School of Natural Sciences, Technische Universit\"at M\"unchen, Am Coulombwall 4, 85748 Garching, Germany
}
\affiliation{ 
Institute of Physics and Astronomy, Technical University Berlin, Hardenbergstrasse 36, 10623 Berlin, Germany
}

\author{H. Ebert}
\affiliation{
Department of Chemistry, Ludwig-Maximilians-Universit\"at M\"unchen,  Butenandtstrasse 5-13, 81377 M\"unchen, Germany
}

\author{E. Zallo}
\email[]{eugenio.zallo@tum.de}
   \affiliation{ 
Walter-Schottky-Institut and TUM School of Natural Sciences, Technische Universit\"at M\"unchen, Am Coulombwall 4, 85748 Garching, Germany
}

\author{J. J. Finley}%
\affiliation{ 
Walter-Schottky-Institut and TUM School of Natural Sciences, Technische Universit\"at M\"unchen, Am Coulombwall 4, 85748 Garching, Germany
}

\date{\today}

\begin{abstract} 

Understanding defects in atomically thin van der Waals (vdW) semiconductors is essential for advancing their use in next-generation optoelectronic and photovoltaic devices. Here, we apply a combination of various impedance spectroscopy techniques to two-dimensional (2D) vdW GaSe doped with silicon (Si) to reconstruct deep trap states across the full bandgap. Deep-level transient spectroscopy reveals three distinct deep states 0.31, 0.88, and 1.40 eV below the conduction band edge. Complementary deep-level optical spectroscopy and photocapacitance measurements identify three deep states at 1.4 and 1.8 eV below the conduction band edge, and 2.0 eV above the valence band edge, with thermal admittance spectroscopy providing additional verification and further resolving two trap states, at 0.16 eV above the valence band edge and at 0.26 eV below the conduction band edge. By comparing the experimentally extracted ionization energies with the predictions of density functional theory, our results attribute these trap states primarily to Si-related defects and metal vacancies. This work presents a comprehensive defect map of Si-doped GaSe, providing critical insights into carrier trapping mechanisms that are essential for optimizing the design of 2D material-based devices for industrial applications.

\end{abstract}

\maketitle

\section{\label{sec:intro}Introduction}

The performance of semiconductor-based (opto)electronic devices is very strongly influenced by deep levels that act as charge traps or recombination centers. These defects, while potentially detrimental to device efficiency, can also serve as functional components in emerging technologies. For example, deep levels can enhance the catalytic activity of 2D materials for the hydrogen evolution reaction, but also operate as spin-active color centers with potential for applications in spin–photon interfaces or for quantum metrology~\cite{Hong_2017,Xiong_2018,Liu_2019,Carbone_review2025}. As the field of two-dimensional (2D) van der Waals (vdW) materials continues to expand, understanding and characterizing electrically active mid-gap defects within this class of semiconductors has become crucial for the advancement of 2D-based devices.
Among the diverse 2D vdW materials, the post-transition metal monochalcogenide (PTMC) GaSe~\cite{cai_synthesis_2019} has garnered significant attention due to its promising electronic and optoelectronic properties. GaSe transistors exhibit high on/off current ratios, exceptional photoresponse, and promising (photo)electrochemical capabilities~\cite{cai_synthesis_2019,yu_review_2024, wang_high-performance_2015,Sorifi_photodet2020,Demissie2024}, making the materials system attractive for next-generation technologies. 

Few-layer GaSe is of particular interest since its valence band lies at the transition between a conventional parabolic dispersion and a caldera-like, ring-shaped dispersion, yielding near-flat bands and a van Hove singularity at the valence band maximum (VBM)~\cite{chen_large-grain_2018, PhysRevB.100.045404, Rybkovskiy2014}. These result in a low carrier group velocity and a strongly enhanced density of states (DOS), which amplifies many-body interactions such as excitonic or correlation effects. These same electronic features, combined with the intrinsically low thermal conductivity, also lead to enhanced Seebeck coefficients, higher power factors, and improved thermoelectric performance~\cite{Cao_GaSe_halfMetallicity2015, Wickramaratne_thermoelectric2015}. For example, theoretical calculations of a GaSe/SnS$_2$ van der Waals heterostructure predict very high thermoelectric performance, with a figure of merit zT reaching 2.99 at 700~K for n-doped GaSe~\cite{Xu_GaSe_SnS22024}. However, the nature of the PTMC band structure also 
makes these materials highly sensitive to disorder: even weak defect potentials can strongly modify carrier transport, thereby degrading mobility, optical response, and thermoelectric performance~\cite{Leykam_flatbanddisorder2013, Das_scattering_ringshaped2019}. \\
Realizing specifically doped GaSe devices, or complementary p-n junctions for photovoltaics, LEDs, and photocatalytic applications, requires reliable n- and p-type doping. However, achieving such doping in atomically thin materials is notoriously challenging due to their high surface‐to‐volume ratio and their strong sensitivity to structural damage. 
In conventional III–V semiconductors such as gallium arsenide, silicon can incorporate on both cation and anion sites, acting either as a donor or an acceptor depending on its lattice position~\cite{Neave_SidopingMBEGaAs1983,Suezawa_SiGaAsPL1991,Dombke_amphotericSiGaAs1996, Ruhstorfer_2020}. This amphoteric behaviour can lead to compensation effects, underscoring the importance of understanding dopant-related defects. In contrast, for GaSe, studies on Si incorporation and its associated defect configurations remain scarce, and understanding these dopant-induced defects is essential for assessing their impact on carrier dynamics and device performance.

Defect characterization in semiconducting 2D materials relies on techniques with varying trade-offs. Photoluminescence, Raman and X-ray photoemission spectroscopies are non-invasive and device-compatible but often provide only qualitative or indirect information\cite{Klein_defects2019,Cançado_GrdefectsRaman2011,Grünleitner_2022}. Scanning tunneling and transmission electron microscopies offer atomic resolution but require complex sample preparation and can be destructive~\cite{Ohta_GaSeSTMdefects2004, Zhou_STEMdefects2013}. In contrast, capacitance-based spectroscopic techniques, such as Deep Level Transient Spectroscopy (DLTS), are a highly sensitive and non-destructive approach to probing electrically active states. DLTS has proven its utility in bulk~\cite{Ci2020} and single-layer~\cite{Zhao2023} transition metal dichalcogenide semiconductors by providing quantitative information on energy levels and capture cross-sections. However, shallow traps and minority carrier bands have not been accessible. Similarly, Deep Level Optical Spectroscopy (DLOS) is a powerful and quantitative approach to accessing deep-level defect states across a wide energy range~\cite{Ghadi2020,app14198785}, enabling the investigation of traps near both band edges.

In this study, we employ a combination of impedance spectroscopy techniques, including DLTS, DLOS, and temperature- and frequency-dependent capacitance-voltage profiling, to investigate defect states in exfoliated Si-doped few-layer GaSe. By systematically exciting GaSe electrically and optically, we reconstruct a comprehensive picture of electrically active defect states within the GaSe bandgap. The experimentally derived trap ionisation energies are then compared with density functional theory (DFT) calculations to identify the specific defects responsible for the observed states. These insights not only elucidate the defect landscape of silicon-doped GaSe and its impact on carrier dynamics but also establish a robust framework for defect characterization in other doped 2D vdW semiconductors.

\section{Results}

\begin{figure*}
\includegraphics[width=\textwidth]{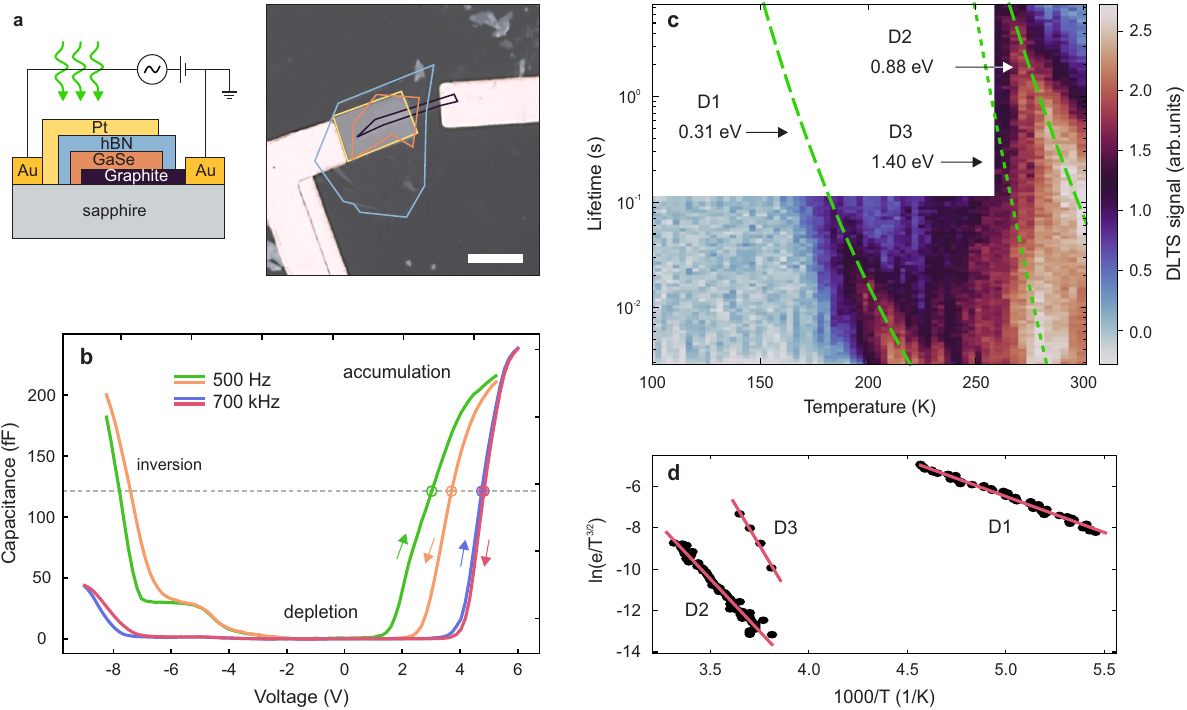}
\caption{\label{DLTS} 
(a) Schematic (left) and optical micrograph (right) of the GaSe metal-insulator-semiconductor (MIS) device (left). (b) Forward and backward CV sweeps at 500~Hz and 700~kHz (room temperature). The grey dotted line indicates the position of half the geometric capacitance. (c) Heat map of the DLTS measurement for temperatures ranging between 100~K and 300~K and (d) Arrhenius plot (black points) of the temperature-dependent lifetime of the identified defects with respective fits (red lines). The dashed green lines in (c) are computed from the Arrhenius fits in (d). To limit measurement duration, temperatures below 257~K were limited to measured lifetimes of a maximum of 100~ms.
}
\end{figure*}

 Modulation of trap occupancy in monolayer 2D semiconductors is challenging due to the atomic-scale thickness, which prevents the formation of a significant vertical depletion width in a Schottky diode, the conventional device geometry for DLTS measurements. By contrast, a metal-insulator-semiconductor (MIS) structure allows for the tuning of the surface Fermi level via a back-gate voltage, enabling the filling and emptying of defects over a wider range of biases. Studies on monolayer MoS$_2$ show DLTS in MIS geometries where trap filling is gate‐modulated, overcoming the limitations of vertical Schottky devices~\cite{Zhao2023}. Moreover, using hBN as the insulator in MIS structures reduces interface state densities substantially due to its atomic flatness and van der Waals bonding~\cite{Knobloch_hBN2021}. Based on this, we prepared the GaSe-based MIS capacitor as depicted in Figure~\ref{DLTS}a. The device stack consists of a rectangular Pt (7~nm) top gate, a 15~nm hBN dielectric, a 7.1~nm GaSe layer, and a thin graphite bottom contact running across the GaSe to achieve a more homogeneous potential distribution. The vdW heterostructure is assembled using a PDMS/PC stamp (propylene carbonate film on polydimethylsiloxane), which is then aligned and transferred onto a sapphire substrate. Afterwards, the Pt gate and electrical contacts are defined and evaporated on top. The sapphire substrate is advantageous for suppressing background-related charges, thereby minimizing parasitic effects, and reducing electrical noise arising from capacitive coupling. 
The doping of GaSe was confirmed by secondary ion mass spectrometry (SIMS), which revealed a Si concentration of $\sim$10$^{20}$~cm$^{-3}$ (see Figure S1 in the Supplementary Information).
Figure~\ref{DLTS}b shows typical room-temperature CV curves recorded at 500~Hz and 700~kHz, by sweeping the back-gate voltage from –8.5~V to +6~V and back, while applying a 50 mV AC signal. At low frequencies, leakage currents restricted the usable gate voltage range to –8 V to +5 V, whereas this effect was suppressed at higher frequencies. As expected for MIS devices, three operating modes can be distinguished in the profile, such as inversion, depletion, and accumulation~\cite{Kurtz2017}. The strong gate-voltage dependence of the capacitance confirms the efficient Fermi-level modulation, and the absence of noticeable frequency dispersion in the strong accumulation regime indicates low series resistance at the contacts. From the CV response, the device clearly exhibits n-type behaviour, consistent with the successful Si incorporation into the GaSe lattice~\cite{Shigetomi_nGaSe-Si2005}. The inversion capacitance is suppressed at high frequencies, as expected, since minority carriers cannot be generated or transported quickly enough to follow the AC modulation. The forward and backward hysteresis, measured at room temperature at half of the geometric capacitance in the weak accumulation regime, is 0.73 and 0.08~V for 500~Hz and 700~kHz, respectively (0.10 and 0.03~V at 70~K), indicating a low density of long-lived traps. In the following sections, further frequency- and temperature-dependent studies will be discussed to extract trap signatures and quantify defect-related states.\\

DLTS measurements on a few-layer GaSe MIS device were performed at the onset of the accumulation regime.  
The populating voltage pulse of 25~ms was set to 4.6~V, and the transients were recorded at 3.8~V. Figure~\ref{DLTS}c shows the heatmap of the DLTS signal constructed from the temperature-dependent capacitance transients (see the Supplementary Information for details on optimization of the measurement and the algorithm used for signal extraction). Each feature in the DLTS signal corresponds to the lifetime of specific mid-gap states as a function of temperature. The emission rate \( e_n \) of an electron deep-level state follows the expression~\cite{Lang_DLTS1974}:

\begin{equation}
e_n = \sigma_n v_{\text{th},n} N_C \exp\left( -\frac{\Delta E}{kT} \right),
\end{equation}

where \( \sigma_n \) is the capture cross-section, $v_{\mathrm{th},n} = \sqrt{\frac{2 k_{B} T}{m^{*}}}$ the thermal velocity of electrons in 2D, $N_{C} = \frac{8\pi\, m^{*}\, k_{B} T}{h^{2}}$ the effective density of states in the 2D conduction band, \( \Delta E \) the activation energy of the trap, \( k \) is the Boltzmann constant, and \( T \) the absolute temperature (see Supplementary Information for further details). An analogous expression holds for hole emission. The activation energies and capture cross-sections of the identified defect states can be determined using a linear fit to the Arrhenius plot after extracting the maxima positions of the DLTS signal for each feature (Figure~\ref{DLTS}d). The DLTS analysis reveals three distinct defect energy levels at 0.31~$\pm$~0.01~eV (D1), 0.88~$\pm$~0.02~eV (D2) and 1.40~$\pm$~0.12~eV (D3) below the conduction band minimum (CBM, E$_c$). By assuming a 2D thermal velocity, 
an effective electron density of states~\cite{Zhao2023} (see Supplementary Information), and employing an in-plane electron effective mass $m_n = 0.2m_0$~\cite{Mooser_1973, Ottaviani_1974}, we determine a capture cross-section for D1 of (7.16~$\pm$~2.51) $\times 10^{-17}~\text{cm}$, consistent with a point defect. The capture cross sections of D2 and D3 could not be extracted due to a large fit error. Note that in the 2D formalism~\cite{Zhao2023} the capture cross section has units of cm rather than cm${^{2}}$.\\

\begin{figure}
\includegraphics[width=0.45\textwidth]{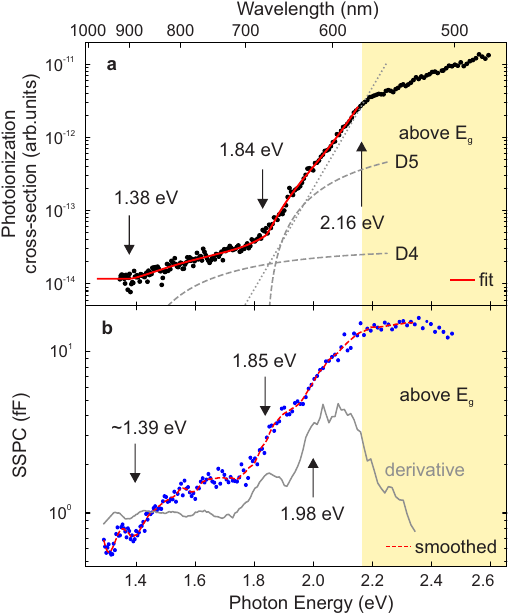}
\caption{\label{DLOS}
(a) Deep Level Optical Spectroscopy (DLOS) and (b) steady-state photocapacitance (SSPC) of the GaSe MIS device. The black dashed lines in (a) show the fitting results using Lucosky's model. The gray line in (b) represents the derivative function used to determine the transitions.
}
\end{figure}

To investigate deep levels near the VBM and construct a clearer picture of the defect landscape, we used DLOS with monochromatic photons with energies $\sim$1.3 to 2.6~eV. DLOS replaces the temperature sweep with an optical source, allowing the detrapping process to be driven by light. This enables probing a broader range of defect states across the band gap, depending on the wavelength used (see Figure S2 and Note S2). Since the measurement records capacitance transients during optical excitation after all defects have been filled with electrons via a voltage pulse, it exclusively probes electron emission from traps relative to the CBM, without contributions from trap-to-trap transitions or optically-driven metastable charge capture by defects. The DLOS spectrum is shown in Figure~\ref{DLOS}a. The data was fitted with two curves based on Lucovsky's model~\cite{Lucovsky_1965} for localized traps, which assumes no lattice relaxation (i.e., neglects the Franck-Condon relaxation energy), and an Urbach-like exponential distribution of traps close to the VBM and CBM. The best fit is characterized by two trap states below the CBM at 1.38~$\pm$~0.02~eV and 1.84~$\pm$~0.01~eV, respectively. The signal increases up to the band edge of few-layer GaSe at 2.16~eV, after which the contribution from band-to-band transitions dominates the spectrum. The exponential fit of the signal close to the band-edge returns an Urbach energy of 63.10~$\pm$~4.90~meV. To validate the DLOS results, we also performed steady-state photocapacitance (SSPC) measurements (see Figure~\ref{DLOS}b), which record the total capacitance change under monochromatic illumination once steady-state conditions are established. Since the steady-state signal reflects a dynamic equilibrium between photo-excitation (detrapping) and optically-driven recapture (retrapping) processes,  the measured capacitance corresponds to the net trap occupancy. A step in the SSPC as a function of excitation energy indicates the activation of a new defect state. SSPC spectroscopy confirms the states close to 1.39~eV and 1.85~eV below the CBM observed in DLOS. The constant slope (see the gray curve in Figure~\ref{DLOS}b) close to the band edge is consistent with an increasing share of defects being excited, compatible with the Urbach-tail seen in the DLOS measurement. An additional feature not evident in the DLOS spectrum appears at 1.98~eV. This discrepancy arises because, as previously mentioned, under the present initialization and transient readout, DLOS selectively probes photo-emission of electrons from pre-filled traps into the conduction band, while SSPC, operating in steady state, also probes charge transfer between traps or from the valence band to traps. We therefore attribute the observed kink at 1.98~eV to a defect state located 0.18~eV below the CBM. Finally, both DLOS and SSPC spectroscopy consistently determine a bandgap of 2.16~eV, in agreement with reported values for GaSe~\cite{FERNELIUS1994275, PhysRevB.100.045404, 10.1063/1.3211967}.\\

\begin{figure*}
\includegraphics[width=\textwidth]{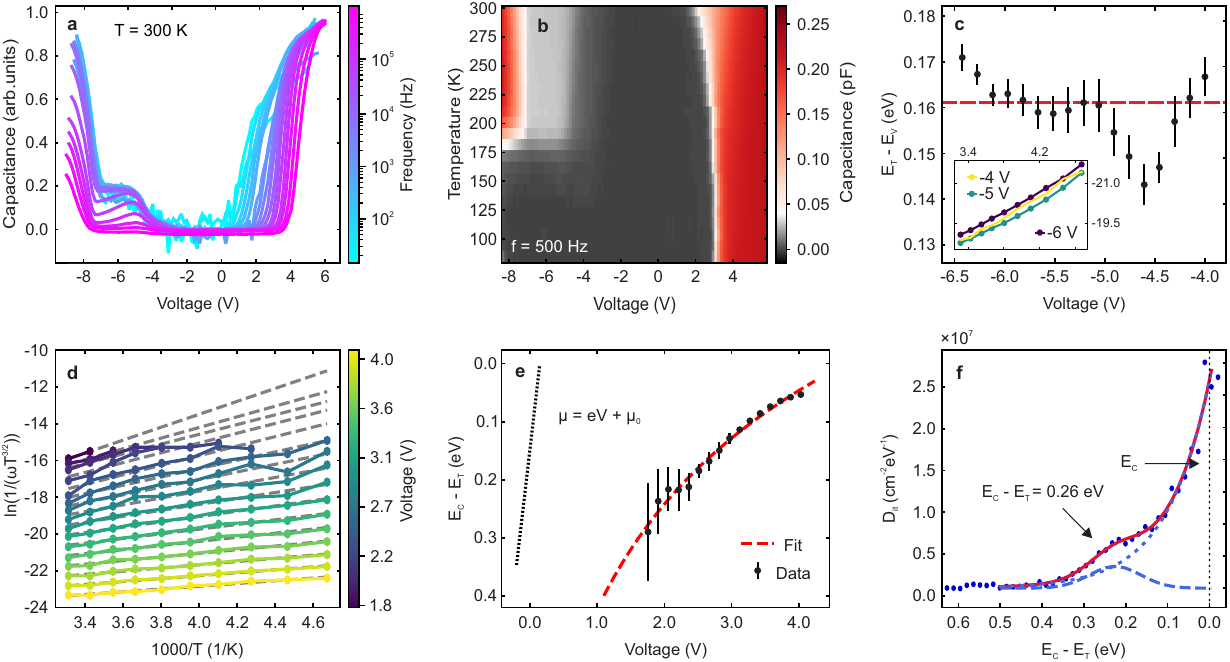}
\caption{\label{TAS} 
(a) CV measurements taken at frequencies ranging from 20~Hz to 700~kHz. (b) Temperature-dependent CV profiling at 500 Hz. (c) Acceptor energy as a function of voltage, as extracted from the Thermal Admittance Spectroscopy (TAS) measurement in the inversion regime. The inset shows the Arrhenius plot of the detrapping lifetime for -4, -5, and -6~V, plotted as $1000/T$ vs.\ $\ln\!\left(1/\omega T^{3/2}\right)$. (d) Arrhenius plots of the detrapping lifetime at various voltages in the accumulation regime. (e) Acceptor energy as a function of voltage, as extracted from (d). The dashed red line is a fit of the voltage-dependent chemical potential $\mu(V)$. The dotted black line indicates $\mu(V)$ for a semiconductor with no gap states. (f) Defect density close to the conduction band. The energy position ($\mu(V)$) is computed from the fit in (e). The red line is a fit using the sum of an exponential function and a Gaussian.
}
\end{figure*}

We continue to investigate the temperature and frequency dependence of the capacitance-voltage (CV) characteristics in detail to probe the dynamic behaviour of the defects. Figure~\ref{TAS}a shows the room-temperature CV curves, measured from -8.5 to 6~V, for a range of frequencies from 20~Hz to 700~kHz. A pronounced frequency dispersion is observed in the weak accumulation regime, which cannot be attributed to series resistance effects, as the strong accumulation regime exhibits negligible frequency dispersion~\cite{Gaur_2019}.
Furthermore, a clear capacitance peak emerges at negative gate voltages but weakens and eventually vanishes as the measurement frequency increases. This is a characteristic signature of trap states with emission rates that are too slow to follow the AC probing signal at higher frequencies~\cite{Sze_Ng_2006,Schroder_2005}.

To elucidate the origin of these trap states, temperature-dependent CV measurements were carried out. Since the emission rate of trap states is thermally activated, changing the temperature shifts the frequency at which these states can respond, resulting in an Arrhenius-type dependence of the cutoff frequency. Figure~\ref{TAS}b presents a false color image of the temperature-dependent CV curves measured at 500~Hz. The capacitance peak at negative voltages vanishes below a certain temperature ($\sim$175~K), consistent with the expected behaviour of discrete trap states whose emission rates fall below the probing frequency. In contrast, the dispersion observed at positive gate voltages does not disappear at a single temperature but rather diminishes gradually over a range of temperatures, depending on the applied bias. This behaviour indicates the presence of a distribution of defect states across a range of energies within the bandgap, such as those associated with an Urbach tail.
It is important to note that the onset of accumulation occurs at lower gate voltages for higher temperatures.

To further investigate the origin of the frequency‐dependent capacitance features and understand the role of minority carriers in the trap emission rates, Thermal Admittance Spectroscopy (TAS) measurements~\cite{SCHIFANO20094344} were performed under different gate bias conditions. TAS yields defect energy levels and capture cross‑sections by tracking the demarcation AC frequency, \(\omega_0\), at which a given defect feature vanishes as temperature changes (see Supplementary Information for details). As the temperature decreases, the defect emission rate, \(1/\tau(T)\), falls toward the probing frequency, \(\omega_{\mathrm{ac}}\). In the limit that  
\[
\omega_0(T) \;=\;\frac{1}{\tau(T)} < \omega_{\mathrm{ac}},
\]
the defect can no longer respond to the AC modulation and ceases to contribute to the total capacitance. By identifying the temperature at which each feature disappears, we reconstruct the temperature dependence of the defect lifetime. An Arrhenius analysis of \(\omega_0(T)\) then allows extraction of the defect activation energy and capture cross‑section, in direct analogy to the DLTS procedure (see Figure~\ref{DLTS}c,d).  
Importantly, both the inversion capacitance and the capacitance peak disappear simultaneously as either the temperature is lowered or the probing frequency is increased, indicating a common, kinetically limited origin. This behavior is consistent with a trap-mediated, thermally activated emission process that supplies minority carriers to the inversion charge.
We exclude a diffusion-limited minority-carrier response, as no systematic dependence of the cutoff frequency on gate voltage is observed. In a diffusion-limited scenario, the characteristic response time would scale with the depletion width and therefore vary with bias. The absence of such a trend instead supports an emission-limited Shockley–Read–Hall process, whose rate is governed by the defect activation energy and capture cross-section.\\

Figure~\ref{TAS}c presents the trap energy, \(E_T-E_v\), as a function of various negative gate voltages. We identify a trap level located at 0.16~$\pm$~0.02~eV above VBM, with a large corresponding capture cross‑section of (2.41~$\pm$~1.24) $\times 10^{-6}~\text{cm}$ (see Supplementary Information for the derivation). Note that for 7.1~nm thick GaSe (8 monolayers), the valence band evolves from a parabolic shape to a caldera‑like, ring-shaped structure, coinciding with the emergence of flat bands and a van Hove singularity near the VBM~\cite{chen_large-grain_2018, PhysRevB.100.045404}. The trap energy and capture cross‑section values reported here are based on the implicit assumption of parabolic bands. As such, they are expected to represent effective values that may deviate significantly from their actual values. Notably, the capture cross‑section should be interpreted with care, since it neglects the effects of low group velocity and high density of states near the van Hove singularity, as well as the non-parabolic character of the effective hole mass ($m_p = -1.9m_0$)~\cite{Aziza_holemeff2018}.\\
The same analysis was applied to the weak‑accumulation regime. Figure~\ref{TAS}d shows the Arrhenius plots of the demarcation frequency, \(\omega_0\), for different gate voltages, and Figure~\ref{TAS}e displays the resulting trap energy relative to the CBM.
We observe that, as the gate voltage increases, the extracted trap energy shifts closer to the CBM, consistent with a continuum of band‑tail states. Consequently, raising the chemical potential enables probing of traps at progressively higher energies, providing an indirect measure of the chemical potential as a function of the applied voltage.

By assuming an exponential band‑tail DOS we obtain for the voltage dependence of the chemical potential $\mu$,
\[
V
=\frac{\mu - \mu_{0}}{e}
+\frac{eD_0E_0}{C_{\rm geom}}
\exp\!\Bigl(\tfrac{\mu}{E_0}\Bigr),
\]  

where $\mu_0$ is the chemical potential at 0~V, $e$ the elementary charge, $D_0$ and $E_0$ the amplitude and characteristic energy of the exponential band-tail, and $C_{\rm geom}$ the geometric capacitance (see Note S4 in the Supplementary Information).
The dashed red line in Figure~\ref{TAS}e shows the fit to this model, while the dotted black line represents the ideal \(\mu(V)\) for a gap without band‑tail states. From the fit, we extract a charge‑neutrality Fermi level position of \(\mu_0 = -0.79~\pm~0.29\ \mathrm{eV}\) relative to the CBM and an Urbach characteristic energy \(E_0 = 0.27~\pm~0.09\ \mathrm{eV}\).  \\
Finally, the interface trap density, \(D_{\mathrm{it}}(V)\), is computed from the measured low‑ and high‑frequency capacitances, \(C_{\mathrm{lf}}(V)\) and \(C_{\mathrm{hf}}(V)\)~\cite{Sze_Ng_2006,Schroder_2005}, using  
\[
D_{\mathrm{it}}
=\frac{1}{q}\Bigl[
\frac{C_{\rm geom}\,C_{\mathrm{lf}}}{C_{\rm geom}-C_{\mathrm{lf}}}
-\frac{C_{\rm geom}\,C_{\mathrm{hf}}}{C_{\rm geom}-C_{\mathrm{hf}}}
\Bigr].
\]  
By mapping \(V\) to energy through the relation for \(\mu(V)\) introduced above, we plot \(D_{\mathrm{it}}(E)\) in Figure~\ref{TAS}f. The trap distribution is well described by the sum of an exponential tail (Urbach‑like) with a characteristic energy of  \(0.10~\pm~0.01\  \mathrm{eV}\) and a Gaussian peak centered at \(0.26~\pm~0.02\ \mathrm{eV}\), indicating a discrete electron trap superimposed on a band‑tail background. These values are in reasonable agreement with the fit in Fig.~\ref{TAS}e, with the remaining difference ascribed to the assumption of 0~K and the omission of the Gaussian peak contribution in the first fit, which, when neglected, artificially broaden the exponential tail.\\

\begin{figure}
\includegraphics[width=0.45\textwidth]{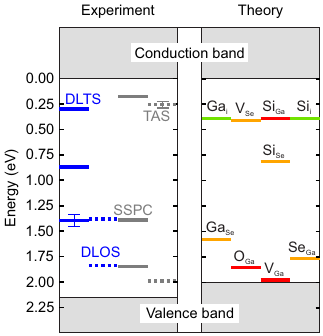}
\caption{\label{theory} 
Comparison between experimentally measured (left) and DFT-calculated (right) mid-gap states and band-edge positions relative to the CBM. Left: Deep levels detected experimentally by DLTS (continuous blue), DLOS (dotted blue), SSPC (continuous gray), and TAS (dotted gray). Right: Defect levels for various point defects calculated by DFT. The colour of each marker indicates the defect site: green for interstitial positions, yellow for selenium sites, and red for gallium sites. Solid black lines indicate the CBM and VBM. 
}
\end{figure}

Our results demonstrate that few-layer GaSe doped with Si hosts multiple energy levels within the band gap. DLTS measurements identified defect states at 0.31~eV, 0.88~eV, and 1.40~eV below the CBM, while TAS revealed a trap at 0.16~eV above the VBM and an Urbach‑like band‑tail density of states near the CBM, together with a another trap at 0.26~eV below the CBM. Optical techniques, including DLOS and SSPC, detected defect levels at 1.4 and 1.8~eV below the CBM, an additional transition at 1.98~eV above the VBM, and the bandgap at 2.16~eV.
While these measurements provide a comprehensive experimental picture of the defect landscape in our materials, they do not yet clarify the microscopic origin of the observed states. Moreover, the influence of silicon doping on these defects remains largely unexplored in the literature. To address these open questions, we performed \textit{ab initio} calculations using density functional theory (DFT) with the HSE06 hybrid functional (see Theoretical Methods in Section~\ref{sec:methods}). These calculations focused on a range of point defects, including gallium and selenium vacancies, oxygen-passivated vacancies, and silicon-related substitutional or interstitial defects. A summary of the calculated defect energy levels is presented in Table~\ref{theorytab} and Figure~\ref{theory}.

\begin{table}[h]
\centering
\caption{\label{theorytab} Calculated defect transition levels relative to band edges.}
\begin{tabular}{lcc}
\hline \hline
 & This work & Literature~\cite{Deák_2020} \\
\hline
Charge transition level & 2.02 eV & 2.26 eV \\
\hline
Ga$_\text{i}$(+/0) & $E_C - 0.39$ & $E_C - 0.2$ \\
V$_\text{Se}$(-/2-) & -- & $E_V + 1.6$ \\
V$_\text{Se}$(0/-) & $E_V + 1.61$ & $E_V + 1.1$ \\
Ga$_\text{Se}$(0/-) & $E_V + 0.45$ & -- \\
Se$_\text{Ga}$(0/-) & $E_V + 0.25$ & -- \\
V$_\text{Ga}$(0/-) & $E_V + 0.04$ & $E_V + 0.1$ \\
Se$_\text{i}$ & inactive & inactive \\
\hline
Si$_\text{Ga}$(+/0) & $E_C - 0.39$ & $E_C - 0.7$ \\
Si$_{\text{i}}$(0/-) & $E_V + 1.63$ & -- \\
Si$_\text{Se}$(0/-) & $E_V + 1.20$ & -- \\
O$_\text{Ga}$(0/-) & $E_V + 0.17$ & -- \\
O$_\text{Se}$ & inactive & -- \\
O$_{\text{i}}$ & inactive & -- \\
\hline \hline
\end{tabular}
\label{tab:defects}
\end{table}


Figure~\ref{theory} highlights the overall good agreement between our DFT calculations and the defect levels observed experimentally, confirming the correspondence between theoretical predictions and DLTS measurements. In particular, the defect levels detected in DLTS align well with those predicted for silicon-related defects. It should be noted that the calculations were performed for bulk GaSe, leading to deviations of up to 0.14~eV (see Figures S4 and S5 in the Supplementary Information) from the experimental values. These deviations are ascribed to confinement-induced changes in the band structure of few-layer GaSe and also the use of plain DFT. Interestingly, the calculations indicate that oxygen-passivated selenium vacancies and interstitial oxygen atoms are electrically inactive.


\section{Discussion}

A defect level at approximately 0.2~eV above the VBM has been consistently reported in GaSe across multiple studies, regardless of the doping type~\cite{https://doi.org/10.1002/pssa.2210380231, Capozzi_1981, PhysRevB.28.4620, PhysRevB.40.3182, 10.1063/1.2831130, Anis_1984, https://doi.org/10.1002/pssa.2210380231, cryst15040372, 10.1063/1.345090}. This defect is commonly attributed to Ga vacancies, which act as acceptor levels. Structural studies using annular dark field-scanning transmission electron microscopy imaging have shown that Ga vacancies are indeed one of the predominant defects in atomically thin GaSe~\cite{doi:10.1021/acsnano.8b08253}. Our TAS measurements reveal a defect level at 0.16~eV above the VBM, in good agreement with the reported values for Ga vacancies. However, this is in contrast with the theoretically obtained values, placing the defect transition level of V$_{\text{Ga}}$ at 0.04~eV above the VBM (see Table~\ref{tab:defects}).\\
The energy calculated by DFT for Si-based defects is in good agreement with our DLTS  and TAS results. The experimentally observed defect levels at 0.31 and 0.88~eV (DLTS), and 0.26~eV (TAS) below the CBM, as well as 1.98~eV (SSPC) above the VBM (which corresponds to a shallow level 0.18~eV below the CBM given the 2.16~eV bandgap), are compatible with Si$_{\text{Ga}}$ and Si$_{\text{i}}$ (both predicted at 0.39~eV below the CBM), which act as an acceptor and donor, respectively, and with Si$_{\text{Se}}$ (predicted at 0.82~eV), which act as an acceptor. However, the applied experimental techniques do not allow unambiguous determination of whether Si occupies Ga sites, interstitial positions, or both. The small offset between the DLTS/TAS and SSPC values is tentatively ascribed to the contribution of a small Franck-Condon relaxation energy, which causes the optical SSPC transition to appear at higher energy than the adiabatic charge-transition level measured by DLTS and TAS~\cite{app14198785}. To assess the likelihood of formation for the different Si-based defects, we computed their formation energy (see Figure S8 in the Supplementary Information) and we found that Si$_{\text{i}}$ exhibits a considerably higher formation energy. In contrast, Si$_{\text{Ga}}$ and Si$_{\text{Se}}$ have comparable and lower formation energies, enabling us to exclude Si$_{\text{i}}$ and assign the observed defect levels to Si$_{\text{Ga}}$ and Si$_{\text{Se}}$. It is worth noting that according to our calculations, interstitial Ga introduces a donor gap state at the same energy as Si$_{\text{Ga}}$ and Si$_{\text{i}}$. Considering the high silicon doping and the presence of Ga vacancies, we attribute the observed state primarily to Si-related defects.\\
Selenium vacancies have previously been associated with acceptor levels in bulk GaSe at approximately 0.5 eV below the CBM~\cite{Micocci_1997,Shigetomi_2007}. Our calculations place V$_{\text{Se}}$ at 0.41~eV below the CBM, overlapping with the energy range of the silicon-related defects. However, given the high Si concentration in our samples with Si$_{\text{Se}}$ exhibiting a formation energy comparable to V$_{\text{Se}}$ and the low formation energy of the electronically inactive O$_{\text{Se}}$ (see Figures S4-S8 in the Supplementary Information), we exclude V$_{\text{Se}}$ as the dominant defect species. This interpretation is consistent with previous structural studies on atomically thin GaSe, which reported the observation of both Se vacancies and O substituting Se~\cite{doi:10.1021/acsnano.8b08253}. Also note that the oxidation of V$_{\text{Se}}$ appears to be the primary process driving the degradation of GaSe under dark conditions~\cite{Shi_oxidation2018}.\\
The defect observed in both DLOS and SSPC spectroscopy, approximately 1.8~eV below the CBM, is consistent with the calculated energy levels of an O-passivated Ga vacancy O$_{\text{Ga}}$ or Se$_{\text{Ga}}$. The former would be in line with the high oxygen content and the presence of Ga vacancies. However, the formation energy for O$_{\text{Ga}}$ is much higher than Se$_{\text{Ga}}$, making it less probable to form. A definitive distinction could be achieved by correlating the DLOS signal with controlled oxygen exposure, which lies beyond the scope of the present work and is left for future studies.
In addition to these point defects, extended defects, such as trigonal defects and metallic inclusions, have been reported in GaSe~\cite{doi:10.1021/acsnano.8b08253, Tonndorf_2017}. While we cannot exclude the presence of large traps, the wide variation in size would lead to a correspondingly broad distribution of defect energetic positions and capture cross-sections.
Finally, the observed Urbach tails at the band edges indicate the presence of shallow, distributed defect states. These are typically associated with defect continua arising from interfacial defects, dielectric defects, or extended defects within the GaSe layers, and their identification is correspondingly more challenging. A comprehensive analysis of these defects is left for future studies.

\section{Conclusions}

In this study, we systematically investigated the defect landscape of few-layer silicon-doped GaSe using a combination of electrical, optical, and temperature-dependent spectroscopies, which reveal defect energies and capture cross-sections consistent with DFT calculations and previous reports. Using these complementary techniques, we identify multiple electrically active defects in silicon-doped GaSe. Specifically, we observe 4 acceptors: three deep levels located at approximately 0.88 eV (Si$_{\mathrm{Se}}$), 1.4 eV (Ga$_{\mathrm{Se}}$, cross-technique average), and 1.8 eV (O$_{\mathrm{Ga}}$/Se$_{\mathrm{Ga}}$, cross-technique average) below the CBM, and one acceptor at $\sim$0.16 eV above the VBM, attributed to V$_{\mathrm{Ga}}$. In addition, we identify a donor state at $\sim$0.25~eV (cross-technique average) below the CBM, consistent with Si$_{\mathrm{Ga}}$. Our results highlight the critical role of silicon in creating states inside the bandgap and suggest that gallium vacancies remain one of the dominant electrically active defects. 

These findings underline the importance of controlling defect chemistry in 2D-GaSe to tailor its electronic properties for next-generation devices. Looking ahead, as scalable, large-area epitaxy of van der Waals PTMCs becomes mature~\cite{zallo_two-dimensional_2023, Bissolo2025, Bissolo2025GaSe, pianetti2025} and device architectures continue to be refined~\cite{Liu2021,Xue2025}, defect control will be a determining factor in achieving reproducible electronic and optoelectronic performance. In this context, the integration of complementary characterization techniques, as demonstrated here, provides a robust framework for systematic optimization of PTMC material properties and represents an essential step toward integration-ready 2D materials technologies.

\section{\label{sec:methods}Experimental Section}

\subsection{Crystal growth}
GaSe bulk crystals were prepared by direct reaction from elements in a quartz ampoule. Gallium (99.9999\%, Wuhan Xinrong New Materials Co., China) and selenium (2--4~mm, 99.9999\%, Wuhan Xinrong New Materials Co., China) corresponding to 25~g of GaSe were placed in a quartz ampoule (25~$\times$~100~mm, quartz purity over 99.99\%) and melt sealed with an oxygen--hydrogen welding torch under high vacuum ($< 1 \times 10^{-3}$~Pa using an oil diffusion pump with LN$_2$ cold trap). The ampoule was placed in a muffle furnace and heated to 900~$^\circ$C using a heating rate of 1~$^\circ$C/min. After 5~hours at that temperature, the furnace was cooled to 500~$^\circ$C using a cooling rate of 5~$^\circ$C/hour and finally freely cooled to room temperature. The ampoule was opened in an argon-filled glovebox.

\subsection{Sample Fabrication}
The few-layer flake (7.1 nm thick, corresponding to $\sim$8 monolayers; see Figure S9 in the Supplementary Information) of gallium selenide was mechanically exfoliated from bulk crystals. The flake was electrically contacted using a few-layer graphite electrode and subsequently transferred onto a c-plane sapphire substrate. To protect the sample from oxidation, a hexagonal boron nitride (hBN) encapsulation layer was used, which also served as the gate dielectric. A 7~nm platinum (Pt) layer was then evaporated onto the top surface to serve as a transparent top electrode.

\subsection{Impedance and Optical Measurements}
 Capacitance-voltage (CV) curves and transient signals were recorded using a Zurich Instruments MFIA Impedance Analyzer, capable of measuring capacitance across frequencies from 1~mHz to 5~MHz and on time scales as short as 10~µs. The samples were mounted on a sample holder in direct contact with the cryostat cold finger to ensure stable thermal conditions. Optical excitation was provided by a NKT Photonics SuperK Fianium laser coupled with a VIS HP8 tunable bandpass filter, offering continuous wavelength tuning from 400~nm to 1000~nm (<2.5~meV bandwidth). Temperature- and time-dependent capacitance measurements were conducted over the range of 75~K to 300~K, with a temperature precision of \(\pm 500~\mathrm{mK}\). Photocapacitance data were acquired at 20~K using monochromatic light with wavelengths ranging from 400~nm to 1000~nm. 


\subsection{Theoretical Methods}
Density functional theory (DFT) ab-initio calculations were employed to identify defects observed via DLTS. Calculations were performed using the Vienna Ab initio Simulation Package (VASP)~\cite{PhysRevB.54.11169} with the projector augmented wave method~\cite{PhysRevB.59.1758}. The cutoff for wave-function expansion was 400~eV. We took van der Waals interactions into account by the Tkatchenko-Scheffler method~\cite{PhysRevLett.102.073005}. Defect states were modelled in a supercell with lattice vectors (4a, 2a+4b, c), where a, b and c are the primitive unit vectors. We took a 2 x 2 x 2 $\Gamma$-centered k-point set. The structure optimization was performed with the exchange-correlation energy functional by Perdew, Burke and Ernzerhof~\cite{PhysRevLett.77.3865} and the total energy calculations were carried out by use of a HSE hybrid functional~\cite{10.1063/1.2404663} with the fraction of exact exchange 0.3. Total energy of charged system  was corrected by the SLABCC code~\cite{FARZALIPOURTABRIZ2019101}. The formation energy was evaluated following the method in Ref.~\cite{RevModPhys.86.253}.
The equilibrium position of Ga$_{\text{i}}$ is between two layers and, on the other hand, Se$_{\text{i}}$ occupies a puckered bond-center interstitial site between a pair of Ga atoms. In the system with V$_{\text{Ga}}$, the pair of the missing Ga atom moves toward the middle of the layer. For other defects,  only small rearrangements of the neighboring atoms were seen. These positions are consistent with other calculations~\cite{Deák_2020}. Si$_{\text{i}}$ and O$_{\text{i}}$ also occupy a puckered bond-center interstitial site between a Ga pair.

\section*{Supporting Information}

 Supporting Information is available from the ...
 Online Library or from the author.

\begin{acknowledgments}
The work was partly funded by the German Research Foundation (DFG) under Germany's Excellence Strategy via the Clusters of Excellence e-conversion (EXC-2089/1-390776260) and the MCQST (EXC-2111-390814868) and via Grants FI 947/7-1, FI 947/8-1 and KO4005-9/1. Z.S. was supported by the ERC-CZ program (project LL2101) from the Ministry of Education, Youth and Sports (MEYS) and by the project Advanced Functional Nanorobots (reg. No. CZ.02.1.01/0.0/0.0/15\_003/0000444 financed by the EFRR).  Z.S.  acknowledges the assistance provided by the Advanced Multiscale Materials for Key Enabling Technologies project, supported by the Ministry of Education Youth, and Sports of the Czech Republic. Project No. CZ.02.01.01/00/22\_008/0004558, co-funded by the European Union.
\end{acknowledgments}

\section*{Conflict of Interest }

The authors declare no conflict of interest

%

\section*{Data Availability Statement}

The data that support the findings of this study are available from the corresponding author upon reasonable request.

\bibliography{DLTS}

@article{Bissolo2025,
  title = {Van der Waals Epitaxy of 2D Gallium Telluride on Graphene: Growth Dynamics and Principal Component Analysis},
  ISSN = {1613-6829},
  url = {http://dx.doi.org/10.1002/smll.202503993},
  DOI = {10.1002/smll.202503993},
  journal = {Small},
  publisher = {Wiley},
  author = {Bissolo,  Michele and Hanke,  Michael and Calarco,  Raffaella and Finley,  Jonathan J. and Koblm\"{u}ller,  Gregor and Lopes,  J. Marcelo J. and Zallo,  Eugenio},
  year = {2025},
  month = may 
}

@article{Bissolo2025GaSe,
  title = {Unveiling the growth mode diagram of GaSe on sapphire},
  volume = {6},
  ISSN = {2662-4443},
  url = {http://dx.doi.org/10.1038/s43246-025-01005-9},
  DOI = {10.1038/s43246-025-01005-9},
  number = {1},
pages = {279},
  journal = {Communications Materials},
  publisher = {Springer Science and Business Media LLC},
  author = {Bissolo,  Michele and Dembecki,  Marco and Belz,  J\"{u}rgen and Schabesberger,  Jan and Bergmann,  Max and Avdienko,  Pavel and Rauscher,  Florian and Ulhe,  Abhilash S. and Riedl,  Hubert and Volz,  Kerstin and Finley,  Jonathan J. and Zallo,  Eugenio and Koblm\"{u}ller,  Gregor},
  year = {2025},
  month = dec 
}

@article{cai_synthesis_2019,
	title = {Synthesis and emerging properties of {2D} layered {III}–{VI} metal chalcogenides},
	volume = {6},
	issn = {1931-9401},
	url = {http://aip.scitation.org/doi/10.1063/1.5123487},
	doi = {10.1063/1.5123487},
	number = {4},
	urldate = {2021-01-28},
	journal = {Applied Physics Reviews},
	author = {Cai, Hui and Gu, Yiyi and Lin, Yu-Chuan and Yu, Yiling and Geohegan, David B. and Xiao, Kai},
	month = dec,
	year = {2019},
	pages = {041312},
}

@article{chen_large-grain_2018,
	title = {Large-grain {MBE}-grown {GaSe} on {GaAs} with a {Mexican} hat-like valence band dispersion},
	volume = {2},
	issn = {2397-7132},
	url = {http://www.nature.com/articles/s41699-017-0047-x},
	doi = {10.1038/s41699-017-0047-x},
	number = {1},
	urldate = {2020-03-16},
	journal = {npj 2D Materials and Applications},
	author = {Chen, Ming-Wei and Kim, HoKwon and Ovchinnikov, Dmitry and Kuc, Agnieszka and Heine, Thomas and Renault, Olivier and Kis, Andras},
	month = dec,
	year = {2018},
	pages = {2},
}

@article{Ci2020,
  title = {Chemical trends of deep levels in van der Waals semiconductors},
  volume = {11},
  ISSN = {2041-1723},
  url = {http://dx.doi.org/10.1038/s41467-020-19247-1},
  DOI = {10.1038/s41467-020-19247-1},
  number = {1},
  journal = {Nature Communications},
  publisher = {Springer Science and Business Media LLC},
  author = {Ci,  Penghong and Tian,  Xuezeng and Kang,  Jun and Salazar,  Anthony and Eriguchi,  Kazutaka and Warkander,  Sorren and Tang,  Kechao and Liu,  Jiaman and Chen,  Yabin and Tongay,  Sefaattin and Walukiewicz,  Wladek and Miao,  Jianwei and Dubon,  Oscar and Wu,  Junqiao},
  year = {2020},
  month = oct 
}

@article{Ghadi2020,
  title = {Influence of growth temperature on defect states throughout the bandgap of MOCVD-grown $\beta$-Ga2O3},
  volume = {117},
  ISSN = {1077-3118},
  url = {http://dx.doi.org/10.1063/5.0025970},
  DOI = {10.1063/5.0025970},
  number = {17},
  journal = {Applied Physics Letters},
  publisher = {AIP Publishing},
  author = {Ghadi,  Hemant and McGlone,  Joe F. and Feng,  Zixuan and Bhuiyan,  A F M Anhar Uddin and Zhao,  Hongping and Arehart,  Aaron R. and Ringel,  Steven A.},
  year = {2020},
  month = oct 
}

@article{Kurtz2017,
  title = {Deep-level spectroscopy in metal–insulator–semiconductor structures},
  volume = {50},
  ISSN = {1361-6463},
  url = {http://dx.doi.org/10.1088/1361-6463/aa5006},
  DOI = {10.1088/1361-6463/aa5006},
  number = {6},
  journal = {Journal of Physics D: Applied Physics},
  publisher = {IOP Publishing},
  author = {Kurtz,  A and Muñoz,  E and Chauveau,  J M and Hierro,  A},
  year = {2017},
  month = jan,
  pages = {065104}
}

@article{wang_high-performance_2015,
	title = {High-performance flexible photodetectors based on {GaTe} nanosheets},
	volume = {7},
	issn = {2040-3364, 2040-3372},
	url = {http://xlink.rsc.org/?DOI=C4NR07313D},
	doi = {10.1039/C4NR07313D},
	number = {16},
	urldate = {2020-03-19},
	journal = {Nanoscale},
	author = {Wang, Zhenxing and Safdar, Muhammad and Mirza, Misbah and Xu, Kai and Wang, Qisheng and Huang, Yun and Wang, Fengmei and Zhan, Xueying and He, Jun},
	year = {2015},
	pages = {7252--7258},
}

@article{yu_review_2024,
	title = {Review of {Nanolayered} {Post}-transition {Metal} {Monochalcogenides}: {Synthesis}, {Properties}, and {Applications}},
	copyright = {https://doi.org/10.15223/policy-029},
	issn = {2574-0970, 2574-0970},
	shorttitle = {Review of {Nanolayered} {Post}-transition {Metal} {Monochalcogenides}},
	url = {https://pubs.acs.org/doi/10.1021/acsanm.3c05984},
	doi = {10.1021/acsanm.3c05984},
	urldate = {2024-08-09},
	journal = {ACS Applied Nano Materials},
	author = {Yu, Mingyu and Hilse, Maria and Zhang, Qihua and Liu, Yongchen and Wang, Zhengtianye and Law, Stephanie},
	month = feb,
	year = {2024},
	pages = {acsanm.3c05984},
}

@article{zallo_two-dimensional_2023,
	title = {Two-dimensional single crystal monoclinic gallium telluride on silicon substrate via transformation of epitaxial hexagonal phase},
	volume = {7},
	issn = {2397-7132},
	url = {https://www.nature.com/articles/s41699-023-00390-4},
	doi = {10.1038/s41699-023-00390-4},
	number = {1},
	urldate = {2023-04-05},
	journal = {npj 2D Materials and Applications},
	author = {Zallo, Eugenio and Pianetti, Andrea and Prikhodko, Alexander S. and Cecchi, Stefano and Zaytseva, Yuliya S. and Giuliani, Alessandro and Kremser, Malte and Borgardt, Nikolai I. and Finley, Jonathan J. and Arciprete, Fabrizio and Palummo, Maurizia and Pulci, Olivia and Calarco, Raffaella},
	month = mar,
	year = {2023},
	pages = {19},
}

@article{Zhao2023,
  title = {Electrical spectroscopy of defect states and their hybridization in monolayer MoS2},
  volume = {14},
  ISSN = {2041-1723},
  url = {http://dx.doi.org/10.1038/s41467-022-35651-1},
  DOI = {10.1038/s41467-022-35651-1},
  number = {1},
  journal = {Nature Communications},
  publisher = {Springer Science and Business Media LLC},
  author = {Zhao,  Yanfei and Tripathi,  Mukesh and Čerņevičs,  Kristiāns and Avsar,  Ahmet and Ji,  Hyun Goo and Gonzalez Marin,  Juan Francisco and Cheon,  Cheol-Yeon and Wang,  Zhenyu and Yazyev,  Oleg V. and Kis,  Andras},
  year = {2023},
  month = jan 
}

@article{Knobloch_hBN2021,
author = {Knobloch, Theresia and Illarionov, Y.Y. and Ducry, Fabian and Schleich, Christian and Wachter, Stefan and Watanabe, Kenji and Taniguchi, Takashi and Mueller, Thomas and Waltl, Michael and Lanza, Mario and Vexler, Mikhail and Luisier, Mathieu and Grasser, Tibor},
year = {2021},
month = {02},
pages = {98-108},
title = {The performance limits of hexagonal boron nitride as an insulator for scaled CMOS devices based on two-dimensional materials},
volume = {4},
journal = {Nature Electronics},
doi = {10.1038/s41928-020-00529-x}
}

@article{Shigetomi_nGaSe-Si2005,
author = {Shigetomi, Shigeru and Ikari, Tetsuo},
year = {2005},
month = {10},
pages = {7521-},
title = {Impurity Effect on Electrical Conduction in n-GaSe Doped with Si, Sn and Ge},
volume = {44},
journal = {Japanese Journal of Applied Physics},
doi = {10.1143/JJAP.44.7521}
}

@article{Lang_DLTS1974,
    author = {Lang, D. V.},
    title = {Deep‐level transient spectroscopy: A new method to characterize traps in semiconductors},
    journal = {Journal of Applied Physics},
    volume = {45},
    number = {7},
    pages = {3023-3032},
    year = {1974},
    month = {07},
    issn = {0021-8979},
    doi = {10.1063/1.1663719},
    url = {https://doi.org/10.1063/1.1663719}
}

@Article{Neave_SidopingMBEGaAs1983,
author={Neave, J. H.
and Dobson, P. J.
and Harris, J. J.
and Dawson, P.
and Joyce, B. A.},
title={Silicon doping of MBE-grown GaAs films},
journal={Applied Physics A},
year={1983},
month={Dec},
day={01},
volume={32},
number={4},
pages={195-200},
issn={1432-0630},
doi={10.1007/BF00820260},
url={https://doi.org/10.1007/BF00820260}
}

@article{Dombke_amphotericSiGaAs1996,
  title = {Microscopic identification of the compensation mechanisms in Si-doped GaAs},
  author = {Domke, C. and Ebert, Ph. and Heinrich, M. and Urban, K.},
  journal = {Phys. Rev. B},
  volume = {54},
  issue = {15},
  pages = {10288--10291},
  numpages = {0},
  year = {1996},
  month = {Oct},
  publisher = {American Physical Society},
  doi = {10.1103/PhysRevB.54.10288},
  url = {https://link.aps.org/doi/10.1103/PhysRevB.54.10288}
}

@article{Suezawa_SiGaAsPL1991,
    author = {Suezawa, M. and Kasuya, A. and Nishina, Y. and Sumino, K.},
    title = {Optical studies of heat‐treated Si‐doped GaAs bulk crystals},
    journal = {Journal of Applied Physics},
    volume = {69},
    number = {3},
    pages = {1618-1624},
    year = {1991},
    month = {02},
    issn = {0021-8979},
    doi = {10.1063/1.347258},
    url = {https://doi.org/10.1063/1.347258}
}

@Article{Grünleitner_2022,
author={Gr{\"u}nleitner, Theresa
and Henning, Alex
and Bissolo, Michele
and Zengerle, Marisa
and Gregoratti, Luca
and Amati, Matteo
and Zeller, Patrick
and Eichhorn, Johanna
and Stier, Andreas V.
and Holleitner, Alexander W.
and Finley, Jonathan J.
and Sharp, Ian D.},
title={Real-Time Investigation of Sulfur Vacancy Generation and Passivation in Monolayer Molybdenum Disulfide via in situ X-ray Photoelectron Spectromicroscopy},
journal={ACS Nano},
year={2022},
month={Dec},
day={27},
publisher={American Chemical Society},
volume={16},
number={12},
pages={20364-20375},
issn={1936-0851},
doi={10.1021/acsnano.2c06317},
url={https://doi.org/10.1021/acsnano.2c06317}
}

@Article{Cançado_GrdefectsRaman2011,
author={Can{\c{c}}ado, L. G.
and Jorio, A.
and Ferreira, E. H. Martins
and Stavale, F.
and Achete, C. A.
and Capaz, R. B.
and Moutinho, M. V. O.
and Lombardo, A.
and Kulmala, T. S.
and Ferrari, A. C.},
title={Quantifying Defects in Graphene via Raman Spectroscopy at Different Excitation Energies},
journal={Nano Letters},
year={2011},
month={Aug},
day={10},
publisher={American Chemical Society},
volume={11},
number={8},
pages={3190-3196},
issn={1530-6984},
doi={10.1021/nl201432g},
url={https://doi.org/10.1021/nl201432g}
}

@Article{Klein_defects2019,
author={Klein, J.
and Lorke, M.
and Florian, M.
and Sigger, F.
and Sigl, L.
and Rey, S.
and Wierzbowski, J.
and Cerne, J.
and M{\"u}ller, K.
and Mitterreiter, E.
and Zimmermann, P.
and Taniguchi, T.
and Watanabe, K.
and Wurstbauer, U.
and Kaniber, M.
and Knap, M.
and Schmidt, R.
and Finley, J. J.
and Holleitner, A. W.},
title={Site-selectively generated photon emitters in monolayer MoS2 via local helium ion irradiation},
journal={Nature Communications},
year={2019},
month={Jun},
day={21},
volume={10},
number={1},
pages={2755},
issn={2041-1723},
doi={10.1038/s41467-019-10632-z},
url={https://doi.org/10.1038/s41467-019-10632-z}
}

@article{Ohta_GaSeSTMdefects2004,
  title = {Atomic structures of defects at GaSe/Si(111) heterointerfaces studied by scanning tunneling microscopy},
  author = {Ohta, Taisuke and Klust, Andreas and Adams, Jonathan A. and Yu, Qiuming and Olmstead, Marjorie A. and Ohuchi, Fumio S.},
  journal = {Phys. Rev. B},
  volume = {69},
  issue = {12},
  pages = {125322},
  numpages = {8},
  year = {2004},
  month = {Mar},
  publisher = {American Physical Society},
  doi = {10.1103/PhysRevB.69.125322},
  url = {https://link.aps.org/doi/10.1103/PhysRevB.69.125322}
}

@Article{Zhou_STEMdefects2013,
author={Zhou, Wu
and Zou, Xiaolong
and Najmaei, Sina
and Liu, Zheng
and Shi, Yumeng
and Kong, Jing
and Lou, Jun
and Ajayan, Pulickel M.
and Yakobson, Boris I.
and Idrobo, Juan-Carlos},
title={Intrinsic Structural Defects in Monolayer Molybdenum Disulfide},
journal={Nano Letters},
year={2013},
month={Jun},
day={12},
publisher={American Chemical Society},
volume={13},
number={6},
pages={2615-2622},
issn={1530-6984},
doi={10.1021/nl4007479},
url={https://doi.org/10.1021/nl4007479}
}

@article{Xiong_2018,
author = {Xiong, Jun and Di, Jun and Xia, Jiexiang and Zhu, Wenshuai and Li, Huaming},
title = {Surface Defect Engineering in 2D Nanomaterials for Photocatalysis},
journal = {Advanced Functional Materials},
volume = {28},
number = {39},
pages = {1801983},
doi = {https://doi.org/10.1002/adfm.201801983},
url = {https://advanced.onlinelibrary.wiley.com/doi/abs/10.1002/adfm.201801983},
eprint = {https://advanced.onlinelibrary.wiley.com/doi/pdf/10.1002/adfm.201801983},
year = {2018}
}

@article{Hong_2017,
author = {Hong, Jinhua and Jin, Chuanhong and Yuan, Jun and Zhang, Ze},
title = {Atomic Defects in Two-Dimensional Materials: From Single-Atom Spectroscopy to Functionalities in Opto-/Electronics, Nanomagnetism, and Catalysis},
journal = {Advanced Materials},
volume = {29},
number = {14},
pages = {1606434},
doi = {https://doi.org/10.1002/adma.201606434},
url = {https://advanced.onlinelibrary.wiley.com/doi/abs/10.1002/adma.201606434},
eprint = {https://advanced.onlinelibrary.wiley.com/doi/pdf/10.1002/adma.201606434},
year = {2017}
}

@Article{Liu_2019,
author={Liu, Xiaolong
and Hersam, Mark C.},
title={2D materials for quantum information science},
journal={Nature Reviews Materials},
year={2019},
month={Oct},
day={01},
volume={4},
number={10},
pages={669-684},
issn={2058-8437},
doi={10.1038/s41578-019-0136-x},
url={https://doi.org/10.1038/s41578-019-0136-x}
}

@article{Sorifi_photodet2020,
author = {Sorifi, Sahin and Moun, Monika and Kaushik, Shuchi and Singh, Rajendra},
title = {High-Temperature Performance of a GaSe Nanosheet-Based Broadband Photodetector},
journal = {ACS Applied Electronic Materials},
volume = {2},
number = {3},
pages = {670-676},
year = {2020},
doi = {10.1021/acsaelm.9b00770},
URL = {https://doi.org/10.1021/acsaelm.9b00770},
eprint = {https://doi.org/10.1021/acsaelm.9b00770}
}

@article{Demissie2024,
        author = {Demissie, Ephrem G. and Siu, Chi-Kit},
        title = {Theoretical Understanding of the Structure–Property Relationship of Oxygen-Doped Gallium Selenide as an Efficient Photocatalyst for Oxygen Evolution Reaction},
        journal = {The Journal of Physical Chemistry C},
        volume = {128},
        number = {25},
        pages = {10397-10406},
        year = {2024},
        doi = {10.1021/acs.jpcc.4c02153},
        URL = {https://doi.org/10.1021/acs.jpcc.4c02153},
        eprint = {https://doi.org/10.1021/acs.jpcc.4c02153}
}

@Article{Mooser_1973,
author={Mooser, E.
and Schl{\"u}ter, M.},
title={The band-gap excitons in gallium selenide},
journal={Il Nuovo Cimento B (1971-1996)},
year={1973},
month={Nov},
day={01},
volume={18},
number={1},
pages={164-208},
issn={1826-9877},
doi={10.1007/BF02832647},
url={https://doi.org/10.1007/BF02832647}
}

@article{Ottaviani_1974,
title = {GaSe: A layer compound with anomalous valence band anisotropy},
journal = {Solid State Communications},
volume = {14},
number = {10},
pages = {933-936},
year = {1974},
issn = {0038-1098},
doi = {https://doi.org/10.1016/0038-1098(74)90396-2},
url = {https://www.sciencedirect.com/science/article/pii/0038109874903962},
author = {G. Ottaviani and C. Canali and F. Nava and Ph. Schmid and E. Mooser and R. Minder and I. Zschokke}
}

@article{Lucovsky_1965,
title = {On the photoionization of deep impurity centers in semiconductors},
journal = {Solid State Communications},
volume = {3},
number = {9},
pages = {299-302},
year = {1965},
issn = {0038-1098},
doi = {https://doi.org/10.1016/0038-1098(65)90039-6},
url = {https://www.sciencedirect.com/science/article/pii/0038109865900396},
author = {G. Lucovsky}
}

@article{PhysRevB.54.11169,
  title = {Efficient iterative schemes for ab initio total-energy calculations using a plane-wave basis set},
  author = {Kresse, G. and Furthm\"uller, J.},
  journal = {Phys. Rev. B},
  volume = {54},
  issue = {16},
  pages = {11169--11186},
  numpages = {0},
  year = {1996},
  month = {Oct},
  publisher = {American Physical Society},
  doi = {10.1103/PhysRevB.54.11169},
  url = {https://link.aps.org/doi/10.1103/PhysRevB.54.11169}
}

@article{PhysRevB.59.1758,
  title = {From ultrasoft pseudopotentials to the projector augmented-wave method},
  author = {Kresse, G. and Joubert, D.},
  journal = {Phys. Rev. B},
  volume = {59},
  issue = {3},
  pages = {1758--1775},
  numpages = {0},
  year = {1999},
  month = {Jan},
  publisher = {American Physical Society},
  doi = {10.1103/PhysRevB.59.1758},
  url = {https://link.aps.org/doi/10.1103/PhysRevB.59.1758}
}

@article{PhysRevLett.102.073005,
  title = {Accurate Molecular Van Der Waals Interactions from Ground-State Electron Density and Free-Atom Reference Data},
  author = {Tkatchenko, Alexandre and Scheffler, Matthias},
  journal = {Phys. Rev. Lett.},
  volume = {102},
  issue = {7},
  pages = {073005},
  numpages = {4},
  year = {2009},
  month = {Feb},
  publisher = {American Physical Society},
  doi = {10.1103/PhysRevLett.102.073005},
  url = {https://link.aps.org/doi/10.1103/PhysRevLett.102.073005}
}

@article{PhysRevLett.77.3865,
  title = {Generalized Gradient Approximation Made Simple},
  author = {Perdew, John P. and Burke, Kieron and Ernzerhof, Matthias},
  journal = {Phys. Rev. Lett.},
  volume = {77},
  issue = {18},
  pages = {3865--3868},
  numpages = {0},
  year = {1996},
  month = {Oct},
  publisher = {American Physical Society},
  doi = {10.1103/PhysRevLett.77.3865},
  url = {https://link.aps.org/doi/10.1103/PhysRevLett.77.3865}
}

@article{10.1063/1.2404663,
    author = {Krukau, Aliaksandr V. and Vydrov, Oleg A. and Izmaylov, Artur F. and Scuseria, Gustavo E.},
    title = {Influence of the exchange screening parameter on the performance of screened hybrid functionals},
    journal = {The Journal of Chemical Physics},
    volume = {125},
    number = {22},
    pages = {224106},
    year = {2006},
    month = {12},
    issn = {0021-9606},
    doi = {10.1063/1.2404663},
    url = {https://doi.org/10.1063/1.2404663},
    eprint = {https://pubs.aip.org/aip/jcp/article-pdf/doi/10.1063/1.2404663/13263224/224106_1_online.pdf},
}

@article{RevModPhys.86.253,
  title = {First-principles calculations for point defects in solids},
  author = {Freysoldt, Christoph and Grabowski, Blazej and Hickel, Tilmann and Neugebauer, J\"org and Kresse, Georg and Janotti, Anderson and Van de Walle, Chris G.},
  journal = {Rev. Mod. Phys.},
  volume = {86},
  issue = {1},
  pages = {253--305},
  numpages = {53},
  year = {2014},
  month = {Mar},
  publisher = {American Physical Society},
  doi = {10.1103/RevModPhys.86.253},
  url = {https://link.aps.org/doi/10.1103/RevModPhys.86.253}
}

@article{Deák_2020,
doi = {10.1088/1361-648X/ab7fdb},
url = {https://doi.org/10.1088/1361-648X/ab7fdb},
year = {2020},
month = {apr},
publisher = {IOP Publishing},
volume = {32},
number = {28},
pages = {285503},
author = {Deák, Peter and Han, Miaomiao and Lorke, Michael and Tabriz, Meisam Farzalipour and Frauenheim, Thomas},
title = {Intrinsic defects of GaSe},
journal = {Journal of Physics: Condensed Matter}
}

@article{FARZALIPOURTABRIZ2019101,
title = {SLABCC: Total energy correction code for charged periodic slab models},
journal = {Computer Physics Communications},
volume = {240},
pages = {101-105},
year = {2019},
issn = {0010-4655},
doi = {https://doi.org/10.1016/j.cpc.2019.02.018},
url = {https://www.sciencedirect.com/science/article/pii/S0010465519300700},
author = {Meisam {Farzalipour Tabriz} and Bálint Aradi and Thomas Frauenheim and Peter Deák}
}

@article{PhysRevB.100.045404,
  title = {Control of excitonic absorption by thickness variation in few-layer GaSe},
  author = {Budweg, Arne and Yadav, Dinesh and Grupp, Alexander and Leitenstorfer, Alfred and Trushin, Maxim and Pauly, Fabian and Brida, Daniele},
  journal = {Phys. Rev. B},
  volume = {100},
  issue = {4},
  pages = {045404},
  numpages = {6},
  year = {2019},
  month = {Jul},
  publisher = {American Physical Society},
  doi = {10.1103/PhysRevB.100.045404},
  url = {https://link.aps.org/doi/10.1103/PhysRevB.100.045404}
}

@article{10.1063/1.3211967,
    author = {Choi, S. G. and Levi, D. H. and Martinez-Tomas, C. and Muñoz Sanjosé, V.},
    title = {Above-bandgap ordinary optical properties of GaSe single crystal},
    journal = {Journal of Applied Physics},
    volume = {106},
    number = {5},
    pages = {053517},
    year = {2009},
    month = {09},
    issn = {0021-8979},
    doi = {10.1063/1.3211967},
    url = {https://doi.org/10.1063/1.3211967},
    eprint = {https://pubs.aip.org/aip/jap/article-pdf/doi/10.1063/1.3211967/14794147/053517_1_online.pdf},
}

@article{FERNELIUS1994275,
title = {Properties of gallium selenide single crystal},
journal = {Progress in Crystal Growth and Characterization of Materials},
volume = {28},
number = {4},
pages = {275-353},
year = {1994},
issn = {0960-8974},
doi = {https://doi.org/10.1016/0960-8974(94)90010-8},
url = {https://www.sciencedirect.com/science/article/pii/0960897494900108},
author = {N.C. Fernelius}
}

@article{Gaur_2019,
doi = {10.1088/2053-1583/ab20fb},
url = {https://doi.org/10.1088/2053-1583/ab20fb},
year = {2019},
month = {may},
publisher = {IOP Publishing},
volume = {6},
number = {3},
pages = {035035},
author = {Gaur, Abhinav and Chiappe, Daniele and Lin, Dennis and Cott, Daire and Asselberghs, Inge and Heyns, Marc and Radu, Iuliana},
title = {Analysis of admittance measurements of MOS capacitors on CVD grown bilayer MoS2},
journal = {2D Materials}
}

@article{doi:10.1021/acsnano.8b08253,
author = {Hopkinson, David
G. and Zólyomi, Viktor and Rooney, Aidan P. and Clark, Nick and Terry, Daniel J. and Hamer, Matthew and Lewis, David J. and Allen, Christopher S. and Kirkland, Angus I. and Andreev, Yuri and Kudrynskyi, Zakhar and Kovalyuk, Zakhar and Patanè, Amalia and Fal’ko, Vladimir I. and Gorbachev, Roman and Haigh, Sarah J.},
title = {Formation and Healing of Defects in Atomically Thin GaSe and InSe},
journal = {ACS Nano},
volume = {13},
number = {5},
pages = {5112-5123},
year = {2019},
doi = {10.1021/acsnano.8b08253},
URL = {   
        https://doi.org/10.1021/acsnano.8b08253
},
eprint = {    
        https://doi.org/10.1021/acsnano.8b08253
}
}

@article{Liu2021,
  title = {Giant Enhancement of Continuous Wave Second Harmonic Generation from Few-Layer GaSe Coupled to High-Q Quasi Bound States in the Continuum},
  volume = {21},
  ISSN = {1530-6992},
  url = {http://dx.doi.org/10.1021/acs.nanolett.1c01975},
  DOI = {10.1021/acs.nanolett.1c01975},
  number = {17},
  journal = {Nano Letters},
  publisher = {American Chemical Society (ACS)},
  author = {Liu,  Zhuojun and Wang,  Jiayi and Chen,  Bo and Wei,  Yuming and Liu,  Wenjing and Liu,  Jin},
  year = {2021},
  month = jul,
  pages = {7405–7410}
}

@article{pianetti2025,
author = {Pianetti, Andrea and Cecchi, Stefano and Hanke, Michael and Finley, Jonathan J. and Arciprete, Fabrizio and Calarco, Raffaella and Zallo, Eugenio},
title = {Epitaxy and Phase Stability of 2D Hexagonal Gallium Telluride on Silicon},
journal = {physica status solidi (RRL) – Rapid Research Letters},
volume = {n/a},
number = {n/a},
year = {2025},
pages = {e202500432},
doi = {https://doi.org/10.1002/pssr.202500432},
url = {https://onlinelibrary.wiley.com/doi/abs/10.1002/pssr.202500432},
eprint = {https://onlinelibrary.wiley.com/doi/pdf/10.1002/pssr.202500432}
}

@misc{Schroder_2005, 
title={Semiconductor Material and Device Characterization}, 
url={http://dx.doi.org/10.1002/0471749095}, 
DOI={10.1002/0471749095}, 
publisher={Wiley}, 
author={Schroder, Dieter K.}, 
year={2005}, 
month=apr
}

@article{SCHIFANO20094344,
title = {Shallow levels in virgin hydrothermally grown n-type ZnO studied by thermal admittance spectroscopy},
journal = {Physica B: Condensed Matter},
volume = {404},
number = {22},
pages = {4344-4348},
year = {2009},
issn = {0921-4526},
doi = {https://doi.org/10.1016/j.physb.2009.09.030},
url = {https://www.sciencedirect.com/science/article/pii/S0921452609010849},
author = {R. Schifano and E.V. Monakhov and B.G. Svensson and W. Mtangi and P. {Janse van Rensburg} and F.D. Auret}
}

@misc{Sze_Ng_2006, 
title={Physics of Semiconductor Devices}, 
url={http://dx.doi.org/10.1002/0470068329}, 
DOI={10.1002/0470068329}, 
publisher={Wiley}, author={Sze, S.M. and Ng, Kwok K.}, 
year={2006}, 
month=apr}

@article{Xue2025,
  title = {Wafer-scale uniform epitaxy of transferable 2D single crystals for gate-all-around nanosheet field effect transistors},
  volume = {16},
  ISSN = {2041-1723},
  url = {http://dx.doi.org/10.1038/s41467-025-65641-y},
  DOI = {10.1038/s41467-025-65641-y},
  number = {1},
  journal = {Nature Communications},
  publisher = {Springer Science and Business Media LLC},
  author = {Xue,  Chengyuan and Tan,  Congwei and Gao,  Xin and Tang,  Junchuan and Sun,  Weiyu and Yin,  Yuling and Wang,  Mengdi and Gao,  Xiaoyin and An,  Hao and Fu,  Boyang and Liu,  Wanqing and Wang,  Yuteng and Li,  Ye and Ding,  Feng and Peng,  Hailin},
  year = {2025},
  month = nov 
}

@article{Cao_GaSe_halfMetallicity2015,
  title = {Tunable Magnetism and Half-Metallicity in Hole-Doped Monolayer GaSe},
  author = {Cao, Ting and Li, Zhenglu and Louie, Steven G.},
  journal = {Phys. Rev. Lett.},
  volume = {114},
  issue = {23},
  pages = {236602},
  numpages = {5},
  year = {2015},
  month = {Jun},
  publisher = {American Physical Society},
  doi = {10.1103/PhysRevLett.114.236602},
  url = {https://link.aps.org/doi/10.1103/PhysRevLett.114.236602}
}

@article{Aziza_holemeff2018,
  title = {Valence band inversion and spin-orbit effects in the electronic structure of monolayer GaSe},
  author = {Ben Aziza, Zeineb and Z\'olyomi, Viktor and Henck, Hugo and Pierucci, Debora and Silly, Mathieu G. and Avila, Jos\'e and Magorrian, Samuel J. and Chaste, Julien and Chen, Chaoyu and Yoon, Mina and Xiao, Kai and Sirotti, Fausto and Asensio, Maria C. and Lhuillier, Emmanuel and Eddrief, Mahmoud and Fal'ko, Vladimir I. and Ouerghi, Abdelkarim},
  journal = {Phys. Rev. B},
  volume = {98},
  issue = {11},
  pages = {115405},
  numpages = {7},
  year = {2018},
  month = {Sep},
  publisher = {American Physical Society},
  doi = {10.1103/PhysRevB.98.115405},
  url = {https://link.aps.org/doi/10.1103/PhysRevB.98.115405}
}

@article{https://doi.org/10.1002/pssa.2210380231,
author = {Manfredotti, C. and Murri, R. and Quirini, A. and Vasanelli, L.},
title = {Photoelectronic properties of n-GaSe},
journal = {physica status solidi (a)},
volume = {38},
number = {2},
pages = {685-693},
doi = {https://doi.org/10.1002/pssa.2210380231},
url = {https://onlinelibrary.wiley.com/doi/abs/10.1002/pssa.2210380231},
eprint = {https://onlinelibrary.wiley.com/doi/pdf/10.1002/pssa.2210380231},
year = {1976}
}

@article{Capozzi_1981,
doi = {10.1088/0022-3719/14/29/021},
url = {https://doi.org/10.1088/0022-3719/14/29/021},
year = {1981},
month = {oct},
publisher = {},
volume = {14},
number = {29},
pages = {4335},
author = {V Capozzi and A Minafra},
title = {Photoluminescence properties of Cu-doped GaSe},
journal = {Journal of Physics C: Solid State Physics}
}

@article{PhysRevB.28.4620,
  title = {Kinetics of radiative recombinations in GaSe and influence of Cu doping on the luminescence spectra},
  author = {Capozzi, Vito},
  journal = {Phys. Rev. B},
  volume = {28},
  issue = {8},
  pages = {4620--4627},
  numpages = {0},
  year = {1983},
  month = {Oct},
  publisher = {American Physical Society},
  doi = {10.1103/PhysRevB.28.4620},
  url = {https://link.aps.org/doi/10.1103/PhysRevB.28.4620}
}

@article{PhysRevB.40.3182,
  title = {Optical spectroscopy of extrinsic recombinations in gallium selenide},
  author = {Capozzi, Vito and Montagna, Maurizio},
  journal = {Phys. Rev. B},
  volume = {40},
  issue = {5},
  pages = {3182--3190},
  numpages = {0},
  year = {1989},
  month = {Aug},
  publisher = {American Physical Society},
  doi = {10.1103/PhysRevB.40.3182},
  url = {https://link.aps.org/doi/10.1103/PhysRevB.40.3182}
}

@article{10.1063/1.2831130,
    author = {Cui, Yunlong and Dupere, Ryan and Burger, Arnold and Johnstone, D. and Mandal, Krishna C. and Payne, S. A.},
    title = {Acceptor levels in GaSe:In crystals investigated by deep-level transient spectroscopy and photoluminescence},
    journal = {Journal of Applied Physics},
    volume = {103},
    number = {1},
    pages = {013710},
    year = {2008},
    month = {01},
    issn = {0021-8979},
    doi = {10.1063/1.2831130},
    url = {https://doi.org/10.1063/1.2831130},
    eprint = {https://pubs.aip.org/aip/jap/article-pdf/doi/10.1063/1.2831130/14818928/013710_1_online.pdf},
}

@article{Anis_1984,
doi = {10.1088/0022-3727/17/6/019},
url = {https://doi.org/10.1088/0022-3727/17/6/019},
year = {1984},
month = {jun},
publisher = {},
volume = {17},
number = {6},
pages = {1229},
author = {M K Anis and A R Piercy},
title = {Electrical conduction in p-GaSe},
journal = {Journal of Physics D: Applied Physics}
}

@Article{cryst15040372,
AUTHOR = {Redkin, Ruslan A. and Onishchenko, Nikolay I. and Kosobutsky, Alexey V. and Brudnyi, Valentin N. and Su, Xinyang and Sarkisov, Sergey Yu.},
TITLE = {Temperature-Dependent Optical Absorption and DLTS Study of As-Grown and Electron-Irradiated GaSe Crystals},
JOURNAL = {Crystals},
VOLUME = {15},
YEAR = {2025},
NUMBER = {4},
ARTICLE-NUMBER = {372},
URL = {https://www.mdpi.com/2073-4352/15/4/372},
ISSN = {2073-4352},
DOI = {10.3390/cryst15040372}
}

@article{10.1063/1.345090,
    author = {Micocci, G. and Siciliano, P. and Tepore, A.},
    title = {Deep level spectroscopy in p‐GaSe single crystals},
    journal = {Journal of Applied Physics},
    volume = {67},
    number = {10},
    pages = {6581-6582},
    year = {1990},
    month = {05},
    issn = {0021-8979},
    doi = {10.1063/1.345090},
    url = {https://doi.org/10.1063/1.345090},
    eprint = {https://pubs.aip.org/aip/jap/article-pdf/67/10/6581/18634629/6581_1_online.pdf},
}

@article{Rybkovskiy2014,
  title = {Transition from parabolic to ring-shaped valence band maximum in few-layer GaS, GaSe, and InSe},
  author = {Rybkovskiy, Dmitry V. and Osadchy, Alexander V. and Obraztsova, Elena D.},
  journal = {Phys. Rev. B},
  volume = {90},
  issue = {23},
  pages = {235302},
  numpages = {9},
  year = {2014},
  month = {Dec},
  publisher = {American Physical Society},
  doi = {10.1103/PhysRevB.90.235302},
  url = {https://link.aps.org/doi/10.1103/PhysRevB.90.235302}
}

@article{Wickramaratne_thermoelectric2015,
    author = {Wickramaratne, Darshana and Zahid, Ferdows and Lake, Roger K.},
    title = {Electronic and thermoelectric properties of van der Waals materials with ring-shaped valence bands},
    journal = {Journal of Applied Physics},
    volume = {118},
    number = {7},
    pages = {075101},
    year = {2015},
    month = {08},
    issn = {0021-8979},
    doi = {10.1063/1.4928559},
    url = {https://doi.org/10.1063/1.4928559},
    eprint = {https://pubs.aip.org/aip/jap/article-pdf/doi/10.1063/1.4928559/14743158/075101_1_online.pdf},
}

@Article{Xu_GaSe_SnS22024,
author ="Xu, Zhiyuan and Xia, Qiong and Zhang, Long and Gao, Guoying",
title  ="A van der Waals p–n heterostructure of GaSe/SnS2: a high thermoelectric figure of merit and strong anisotropy",
journal  ="Nanoscale",
year  ="2024",
volume  ="16",
issue  ="5",
pages  ="2513-2521",
publisher  ="The Royal Society of Chemistry",
doi  ="10.1039/D3NR05284B",
url  ="http://dx.doi.org/10.1039/D3NR05284B"}

@article{Leykam_flatbanddisorder2013,
  title = {Flat band states: Disorder and nonlinearity},
  author = {Leykam, Daniel and Flach, Sergej and Bahat-Treidel, Omri and Desyatnikov, Anton S.},
  journal = {Phys. Rev. B},
  volume = {88},
  issue = {22},
  pages = {224203},
  numpages = {6},
  year = {2013},
  month = {Dec},
  publisher = {American Physical Society},
  doi = {10.1103/PhysRevB.88.224203},
  url = {https://link.aps.org/doi/10.1103/PhysRevB.88.224203}
}

@article{Das_scattering_ringshaped2019,
  title = {Charged impurity scattering in two-dimensional materials with ring-shaped valence bands: GaS, GaSe, InS, and InSe},
  author = {Das, Protik and Wickramaratne, Darshana and Debnath, Bishwajit and Yin, Gen and Lake, Roger K.},
  journal = {Phys. Rev. B},
  volume = {99},
  issue = {8},
  pages = {085409},
  numpages = {10},
  year = {2019},
  month = {Feb},
  publisher = {American Physical Society},
  doi = {10.1103/PhysRevB.99.085409},
  url = {https://link.aps.org/doi/10.1103/PhysRevB.99.085409}
}

@article{Micocci_1997,
author = {Micocci, G. and Serra, A. and Tepore, A.},
title = {Impurity Levels in Sn-Doped GaSe Semiconductor},
journal = {physica status solidi (a)},
volume = {162},
number = {2},
pages = {649-659},
doi = {https://doi.org/10.1002/1521-396X(199708)162:2<649::AID-PSSA649>3.0.CO;2-Z},
url = {https://onlinelibrary.wiley.com/doi/abs/10.1002/1521-396X%28199708%29162%3A2%3C649%3A%3AAID-PSSA649%3E3.0.CO%3B2-Z},
eprint = {https://onlinelibrary.wiley.com/doi/pdf/10.1002/1521-396X%28199708%29162%3A2%3C649%3A%3AAID-PSSA649%3E3.0.CO%3B2-Z},
year = {1997}
}

@article{Shigetomi_2007,
doi = {10.1143/JJAP.46.5774},
url = {https://doi.org/10.1143/JJAP.46.5774},
year = {2007},
month = {sep},
publisher = {},
volume = {46},
number = {9R},
pages = {5774},
author = {Shigetomi, Shigeru and Ikari, Tetsuo},
title = {Optical and Electrical Properties of p-GaSe Doped with In},
journal = {Japanese Journal of Applied Physics}
}

@Article{Shi_oxidation2018,
author ="Shi, Li and Li, Qiang and Ouyang, Yixin and Wang, Jinlan",
title  ="Effect of illumination and Se vacancies on fast oxidation of ultrathin gallium selenide",
journal  ="Nanoscale",
year  ="2018",
volume  ="10",
issue  ="25",
pages  ="12180-12186",
publisher  ="The Royal Society of Chemistry",
doi  ="10.1039/C8NR01533C",
url  ="http://dx.doi.org/10.1039/C8NR01533C"}

@article{Tonndorf_2017,
doi = {10.1088/2053-1583/aa525b},
url = {https://doi.org/10.1088/2053-1583/aa525b},
year = {2017},
month = {feb},
publisher = {IOP Publishing},
volume = {4},
number = {2},
pages = {021010},
author = {Tonndorf, Philipp and Schwarz, Stefan and Kern, Johannes and Niehues, Iris and Del Pozo-Zamudio, Osvaldo and Dmitriev, Alexander I and Bakhtinov, Anatoly P and Borisenko, Dmitry N and Kolesnikov, Nikolai N and Tartakovskii, Alexander I and Michaelis de Vasconcellos, Steffen and Bratschitsch, Rudolf},
title = {Single-photon emitters in GaSe},
journal = {2D Materials}
}

@article{Ruhstorfer_2020,
    author = {Ruhstorfer, Daniel and Mejia, Simon and Ramsteiner, Manfred and Döblinger, Markus and Riedl, Hubert and Finley, Jonathan J. and Koblmüller, Gregor},
    title = {Demonstration of n-type behavior in catalyst-free Si-doped GaAs nanowires grown by molecular beam epitaxy},
    journal = {Applied Physics Letters},
    volume = {116},
    number = {5},
    pages = {052101},
    year = {2020},
    month = {02},
    issn = {0003-6951},
    doi = {10.1063/1.5134687},
    url = {https://doi.org/10.1063/1.5134687},
    eprint = {https://pubs.aip.org/aip/apl/article-pdf/doi/10.1063/1.5134687/14529781/052101_1_online.pdf},
}

@article{Carbone_review2025,
    author = {Carbone, Amedeo and Bendixen-Fernex de Mongex, Diane-Pernille and Krasheninnikov, Arkady V. and Wubs, Martijn and Huck, Alexander and Hansen, Thomas W. and Holleitner, Alexander W. and Stenger, Nicolas and Kastl, Christoph},
    title = {Creation and microscopic origins of single-photon emitters in transition-metal dichalcogenides and hexagonal boron nitride},
    journal = {Applied Physics Reviews},
    volume = {12},
    number = {3},
    pages = {031333},
    year = {2025},
    month = {09},
    issn = {1931-9401},
    doi = {10.1063/5.0278132},
    url = {https://doi.org/10.1063/5.0278132},
    eprint = {https://pubs.aip.org/aip/apr/article-pdf/doi/10.1063/5.0278132/20706884/031333_1_5.0278132.pdf},
}

@Article{app14198785,
AUTHOR = {Kruszewski, Piotr and Sakowski, Konrad and Gościński, Krzysztof and Prystawko, Paweł},
TITLE = {The Photoionization Processes of Deep Trap Levels in n-GaN Films Grown by MOVPE Technique on Ammono-GaN Substrates},
JOURNAL = {Applied Sciences},
VOLUME = {14},
YEAR = {2024},
NUMBER = {19},
ARTICLE-NUMBER = {8785},
URL = {https://www.mdpi.com/2076-3417/14/19/8785},
ISSN = {2076-3417},
DOI = {10.3390/app14198785}
}

\end{document}